\documentclass{aa}
\usepackage{graphics}
\usepackage{graphics}

\def\spose#1{\hbox to 0pt{#1\hss}}
\def\gsim{\mathrel{\spose{\lower 3pt\hbox{$\mathchar"218$}}
          \raise 2.0pt\hbox{$\mathchar"13E$}}}
\def\lsim{\mathrel{\spose{\lower 3pt\hbox{$\mathchar"218$}}
          \raise 2.0pt\hbox{$\mathchar"13C$}}}

\begin{document}

\title{Afterglow calculation in the electromagnetic
model for gamma-ray bursts}

\author{F. Genet, F. Daigne \& R. Mochkovitch}

\institute{Institut d'Astrophysique de Paris - UMR 7095 CNRS et Universit\'e
Pierre et Marie Curie, \\
98 bis, boulevard Arago, 75014 Paris, France\\ 
\tt e-mail: genet@iap.fr}
\titlerunning{Afterglow calculation in the Lyutikov-Blandford model}
\abstract
{}
{We compute the afterglow of gamma-ray bursts 
produced by purely electromagnetic outflows to see if it shows characteristic 
signatures
differing from those obtained with the standard 
internal/external shock model.}
{Using a simple approach for the injection of electromagnetic
energy to the forward shock we obtain the afterglow evolution both during
the period of activity of the central source and after. Our method 
equally applies
to a variable source.}
{Afterglow light curves in the visible and X-ray bands are computed both
for a uniform medium and a stellar wind environment. 
They are brighter at early times than afterglows  
obtained with the internal/external shock model but relying only on these
differences to discriminate between models is not sufficient.} 
{}

\authorrunning{F. Genet et al}
\titlerunning{Afterglow in the electromagnetic model for GRBs}

\maketitle

\section{Introduction}
Lyutikov and Blandford (2003) proposed an alternative
to the standard fireball model where the central engine produces 
a purely electromagnetic outflow instead of a  
re\-la\-tivistic baryonic wind.
Observationally this electro\-ma\-gne\-tic model (hereafter EMM) differs 
from the standard internal/external shock model by the absence of 
any reverse shock contribution, a different early afterglow evolution
and a high polarization of the prompt emission (Lyutikov, 2004).  
In this paper we concentrate on the early afterglow (while the central
source is still active) and compare the EMM to the standard model
in X-rays and the visible for a uniform external medium or a 
wind environment. In Sect.2 we obtain simple equations that go\-vern the evolution
of the forward shock propagating in the burst environment. Their solutions 
are used in Sect.3 to compute afterglow light curves which are compared to 
those obtained in the standard model for the same total injected energy. 
We discuss our results in Sect.4 and conclude that it likely will be 
difficult to decide between models from afterglow observations only.
\section{Dynamics of the forward shock}
\begin{figure*}{}
\begin{center}
\begin{tabular}{cc}
\resizebox{0.48\hsize}{!}{\includegraphics{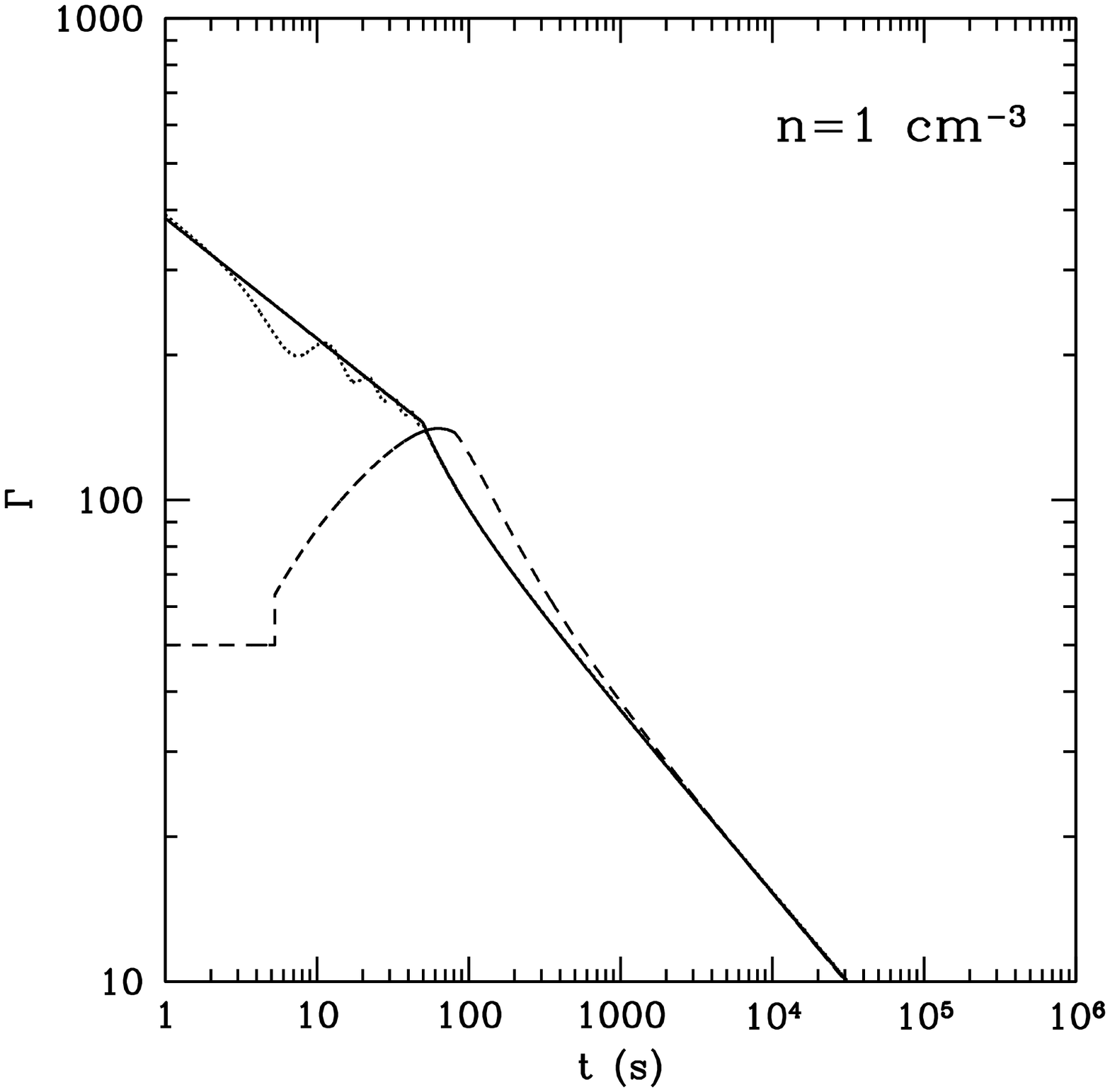}}&
\resizebox{0.48\hsize}{!}{\includegraphics{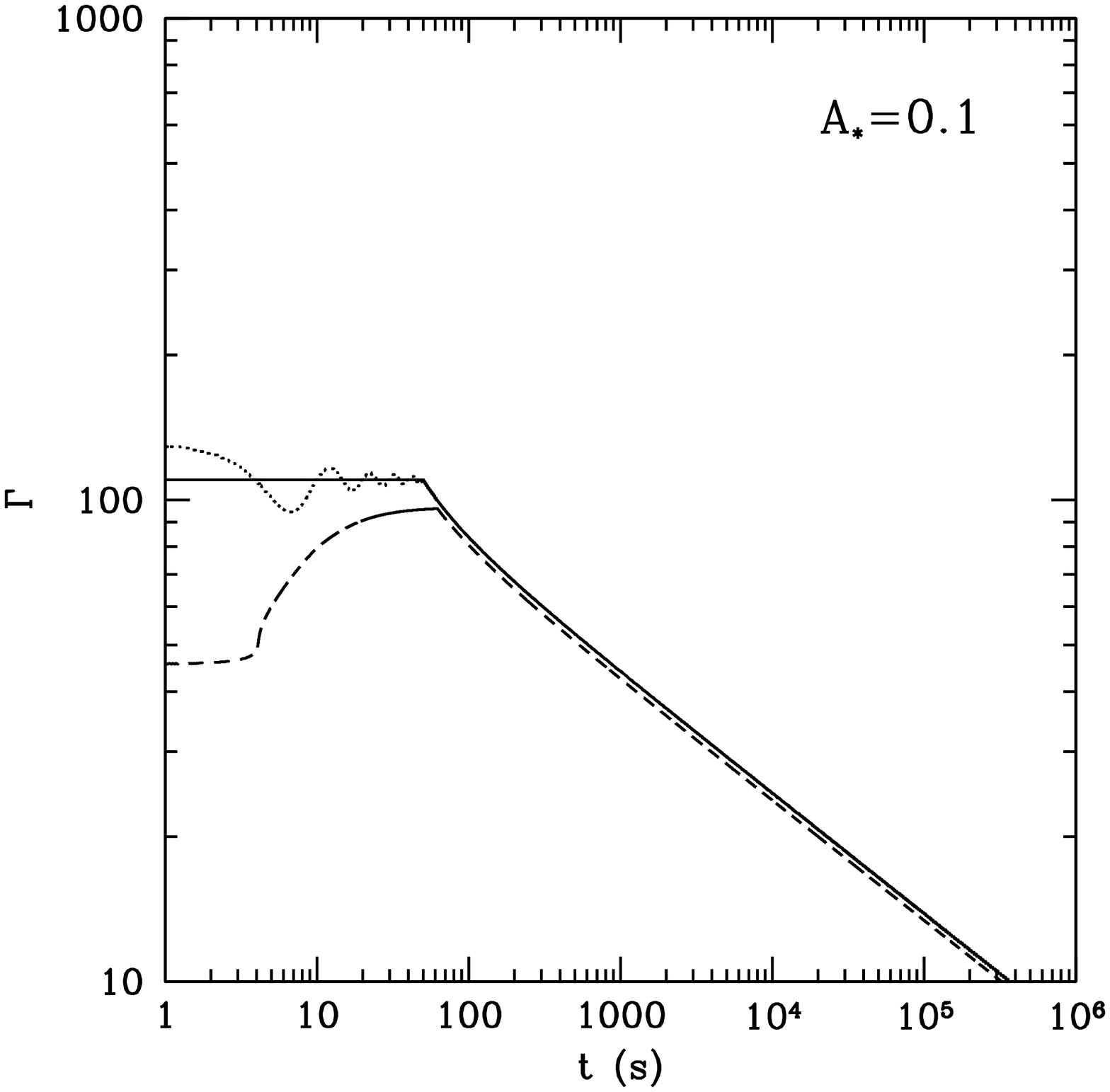}}\\
\end{tabular}
\end{center}
\caption{Evolution of the Lorentz factor in the different models considered;
left panel: uniform external medium of density $n=1$ cm$^{-3}$; right panel:
stellar wind with $A_*=A/(5\,10^{11}$ g.cm$^{-1}$) = 0.1. The full, dotted and dashed lines
respectively correspond to the EMM with a constant
$L_{\rm EM}=10^{52}$ erg.s$^{-1}$, a variable $L_{\rm EM}(t)$ given by Eq.(13) 
and to the standard
internal/external shock model with $\dot E=10^{52}$ erg.s$^{-1}$ and an initial distribution of the
Lorentz factor in the relativistic wind given by Eq.(14).
}
\end{figure*}  
In the context of the EMM, electromagnetic energy released by the central source 
directly leads to the formation of a forward shock propagating 
in the external medium. 
We obtain the evolution of this forward shock by wri\-ting the 
conservation of energy-momentum of the swept-up mass as it accumulates
electromagnetic energy 
\begin{equation}
\begin{array}{l}
E_{\rm EM}+Mc^2=M\Gamma\Gamma_i c^2 \\
E_{\rm EM}=M\beta\Gamma\Gamma_i c^2
\end{array}
\end{equation}
where $M$ is the swept-up mass, $\beta$ and $\Gamma$ the velocity and
Lorentz factor for the bulk motion of the shocked material 
and $\Gamma_i$ the Lorentz
factor for internal motions. The energy $E_{\rm EM}$ received from the source
reads 
\begin{equation}
E_{\rm EM}=\int_0^{t_{s}} L_{\rm EM}(t)dt
\end{equation}
where $L_{\rm EM}(t)$ is the source electromagnetic power
and $t_{s}$ the time in the source frame  
(which is also the observer
time modulo the $(1+z)$ factor for time dilation). 

This can be easily understood if we define the time $t_e$ when the shocked 
material is located at radius $R$. The emitted afterglow signal 
will reach the observer at 
\begin{equation}
t_{\rm obs}=t_e+{(D-R)\over c}\,
\end{equation} 
where $D$ is the distance of the source to the observer.
At time $t_e$, all power produced by the source up to a time $t_s$ 
such as
\begin{equation}   
c(t_e-t_s)=R
\end{equation} 
has been received by the moving shell and Eq.(3) and (4) then lead to
\begin{equation}   
t_{\rm obs}=t_s+{D\over c}\,. 
\end{equation}
Going back to Eq.(1) and
after eli\-mi\-na\-tion
of $\Gamma_i$  we get
\begin{equation}
2Mc^2 \Gamma^2 \sim E_{\rm EM}
\end{equation}
this approximation being valid in the relativistic phase ($\Gamma\gg 1$).
Differentiation of Eq.(6) with respect to observer time 
(with the
simplified notation $t=t_{\rm obs}$) yields
\begin{equation}
L_{\rm EM}(t)={16\pi A c^2\over 3-s}\left(R^{\,3-s}\,\Gamma {d\Gamma\over dt}+
(3-s) \,c 
R^{\,2-s}\Gamma^4\right)
\end{equation}
where we have written 
\begin{equation}
M(R)={4\pi A\over 3-s} R^{\,3-s}
\end{equation}
with $A=\rho$ and $s=0$ for a uniform medium of density 
$\rho$ and $A=5\,10^{11}A_*$ g.cm$^{-1}$ and $s=2$ for a wind environment. 
With the additional relation between observer time and
shock radius
\begin{equation}
{dR\over dt}=2c \Gamma^2
\end{equation}
the problem can be solved for any law
$L_{\rm EM}(t)$. With a constant $L_{\rm EM}$ it can be easily shown that the 
solutions of Eq.(7) and (9) are
\begin{equation}
\Gamma=\left(Q/2\right)^{1/2} (ct)^{-1/4}\ \ \ \ {\rm and}\ \ \ \ 
R=2Q(ct)^{1/2}
\end{equation}
with 
\begin{equation}
Q=\left({3L_{\rm EM}\over 32\pi\rho c^3}\right)^{1/4}
\end{equation}
for a uniform medium and
\begin{equation}
\Gamma=\left({L_{\rm EM}\over 16\pi A c^3}\right)^{1/4}\ \ \ \ {\rm and}\ \ \ \ 
R=2c \Gamma^2 t
\end{equation}
for a stellar wind. With a variable $L_{\rm EM}(t)$, Eq.(7) and (9) are 
integrated
using Eq.(10) or (12) as initial conditions with $L_{\rm EM}=
L_{\rm{EM}}(t=0)$. 
Figure 1 shows the resulting $\Gamma(t)$ for a constant $L_{\rm EM}=10^{52}$ 
erg.s$^{-1}$ or
$L_{\rm {EM}}(t)$ given by
\begin{equation}
L_{\rm EM}(t)=10^{52}\left[1+{\rm cos}2\pi\,\left({t\over 10\,{\rm s}}\right)\right]
\ \ \ {\rm erg.s}^{-1}\ .
\end{equation} 
In both cases the source is supposed to be active for 50 s.
For comparison we also plot in Fig.1 the evolution of the
Lorentz factor  computed 
in the standard internal/external shock model. In that case, we 
assume that a constant
kinetic power $\dot E=10^{52}$ erg.s$^{-1}$ is injected in the relativistic
wind for a duration $t_{\rm W}=50$ s. 
We consider a single pulse burst 
produced by a distribution of the Lorentz factor of the form
\begin{equation}
\Gamma(t)={\Gamma_{\rm max}+\Gamma_{\rm min}\over 2}
-{\Gamma_{\rm max}-\Gamma_{\rm min}\over 2}\,{\rm cos}\,\Big(\pi {t\over
0.2\,t_{\rm W}}\Big)
\end{equation}
if $t< 0.2\,t_{\rm W}$ and $\Gamma(t)=\Gamma_{\rm max}$ if
$t> 0.2\,t_{\rm W}$; $\Gamma_{\rm max}=200$ and $\Gamma_{\rm min}=50$
are the maximum and minimum values of the Lorentz factor so that the most 
rapid part of the wind is decelerated by the slower one which was 
emitted previously.
To follow the wind evolution and its interaction with the external medium
we represent the flow with a large number ($\gsim 1000$) of discrete shells 
which interact 
by direct collision only, pressure waves being neglected (Daigne and
Mochkovitch, 1998). The action 
of the external medium is included by progressively adding the swept-up
mass to the front shell of the wind.
In all cases, the Lorentz factor relaxes to the 
standard Blandford-McKee (1976) solution after a few $10^2$ s. 
However at early times $\Gamma$ is larger in the EMM,
especially for a uniform external medium.
\begin{figure*}{}
\begin{center}
\begin{tabular}{cc}
\resizebox{0.48\hsize}{!}{\includegraphics{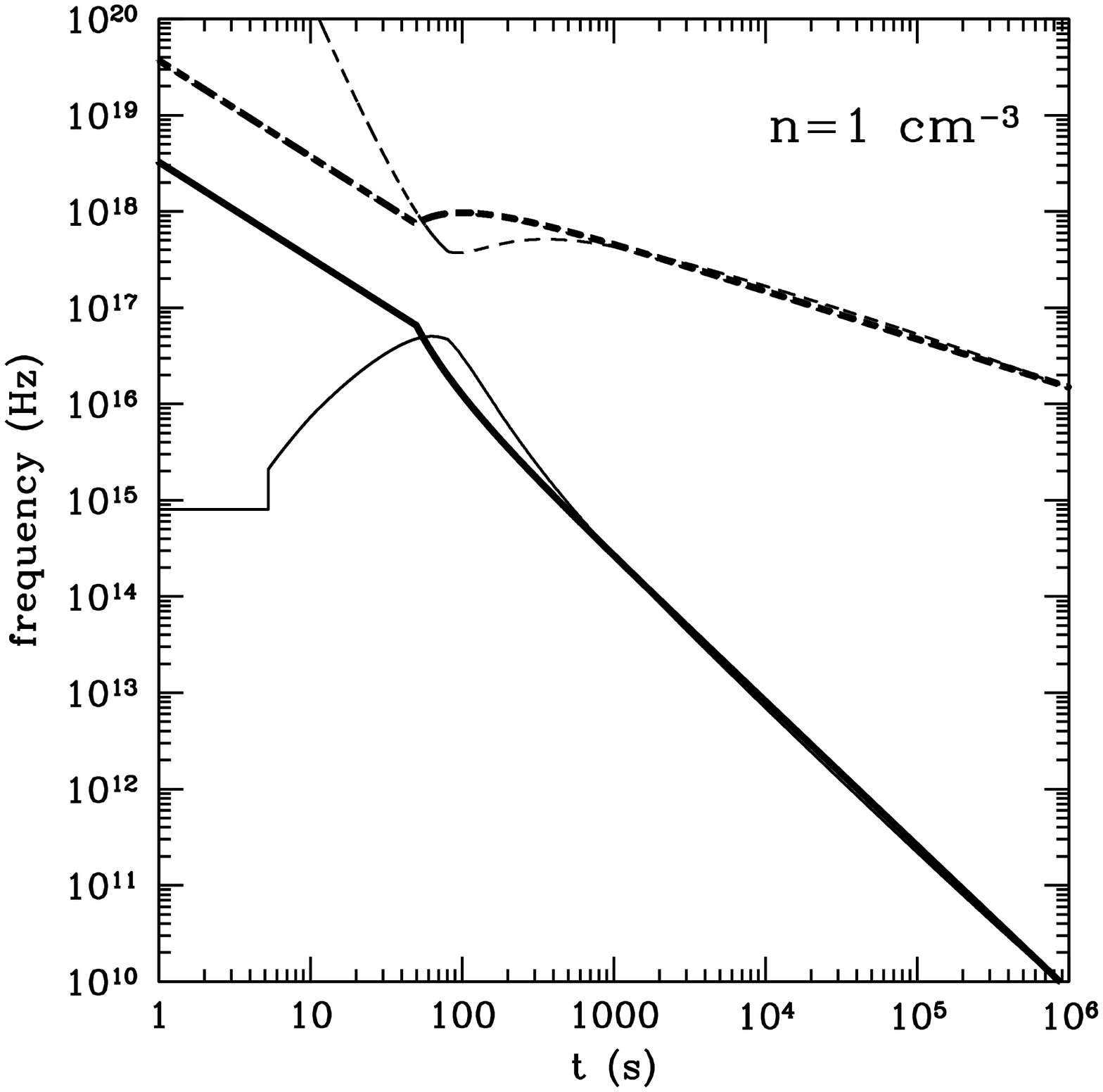}}&
\resizebox{0.48\hsize}{!}{\includegraphics{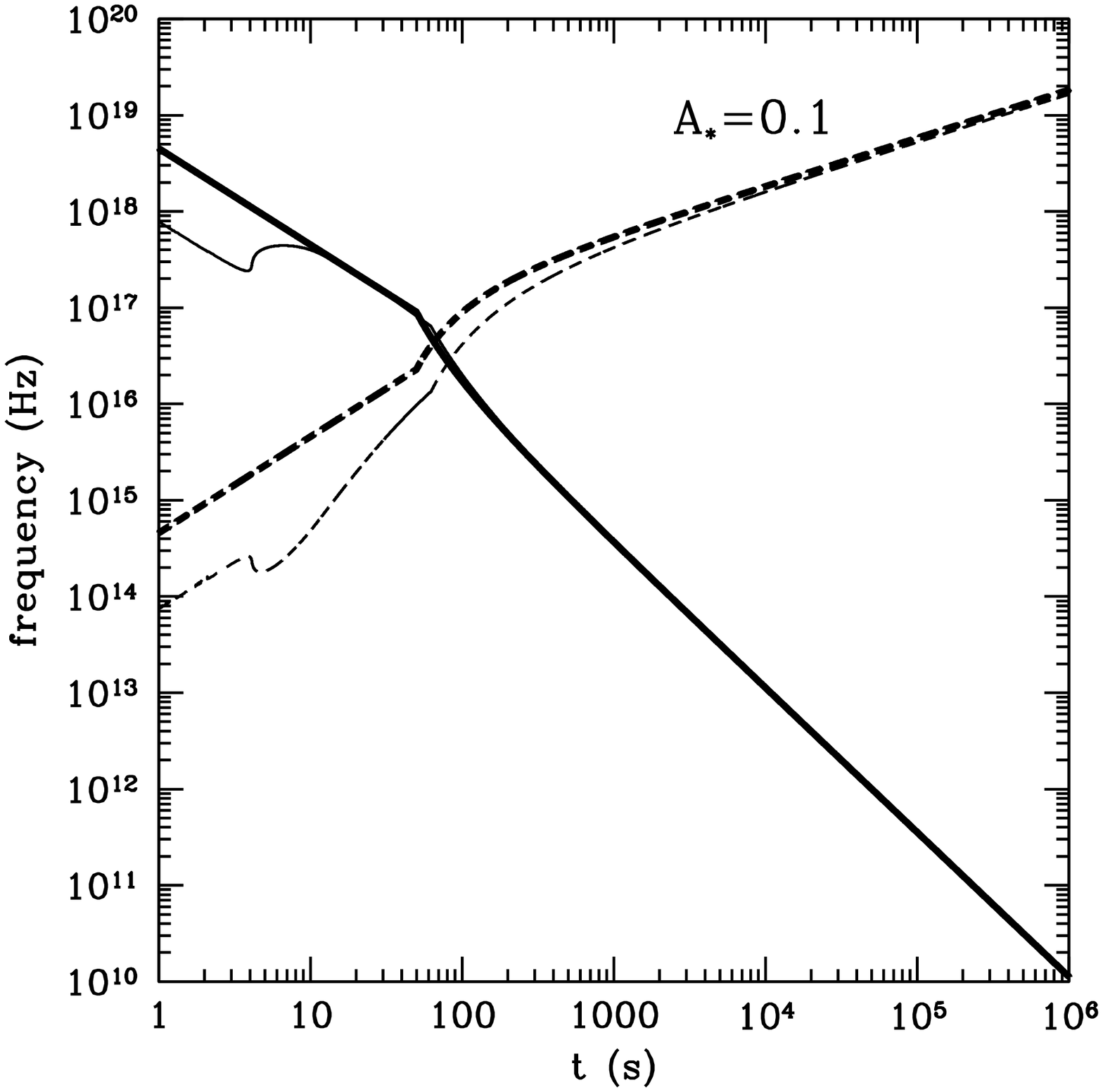}}\\
\end{tabular}
\end{center}
\caption{The two frequencies $\nu_m$ (full lines) and $\nu_c$ (dashed lines)
in the EMM (thick lines) and the standard model (thin lines);
left panel: uniform external medium of density $n=1$ cm$^{-3}$; right panel:
stellar wind with $A_*=0.1$.
}
\end{figure*}  
\section{Afterglow calculation}
\subsection{Method}
Using the dynamical evolution of the forward shock obtained above
we have calculated afterglow lightcurves in X-rays and in the V band
for both the
electromagnetic and the standard model.  
We obtain
the two critical frequencies $\nu_m$ and $\nu_c$ following Sari, Piran and 
Narayan (1998). 
They are represented in Fig.2 for a uniform medium and a stellar 
wind environment.
In the EMM with a uniform medium and a constant source, 
the ratio $\nu_m/\nu_c$ 
is given by
\begin{equation}
{\nu_m\over \nu_c}=0.065 \,\epsilon_{e,-1}^2\epsilon_{B,-3}^2\,n\,L_{52}
\end{equation}
where $n$ is the density in cm$^{-3}$, $L_{52}$ the electro\-ma\-gne\-tic power 
in units 
of $10^{52}$ erg.s$^{-1}$, $\epsilon_{e,-1}=\epsilon_{e}/10^{-1}$ 
and $\epsilon_{B,-3}=\epsilon_{B}/10^{-3}$ (the numerical factor in Eq.(15)
corresponding to the electron power law index $p=2.5$). 
The cooling
regime therefore remains fixed as long as the source is active (and does not vary).
In Fig.2 we have adopted    
$\epsilon_{e,-1}=\epsilon_{B,-3}=L_{52}=1$ and $n=1$ or $A_*=0.1$
in the uniform medium or wind case respectively.

With these parameters the afterglow is always in the slow cooling
regime in the uniform medium case while in the wind case
it is in fast cooling
while the source is active and moves to slow cooling  
shortly after. 
\begin{figure*}{}
\begin{center}
\begin{tabular}{cc}
\resizebox{0.48\hsize}{!}{\includegraphics{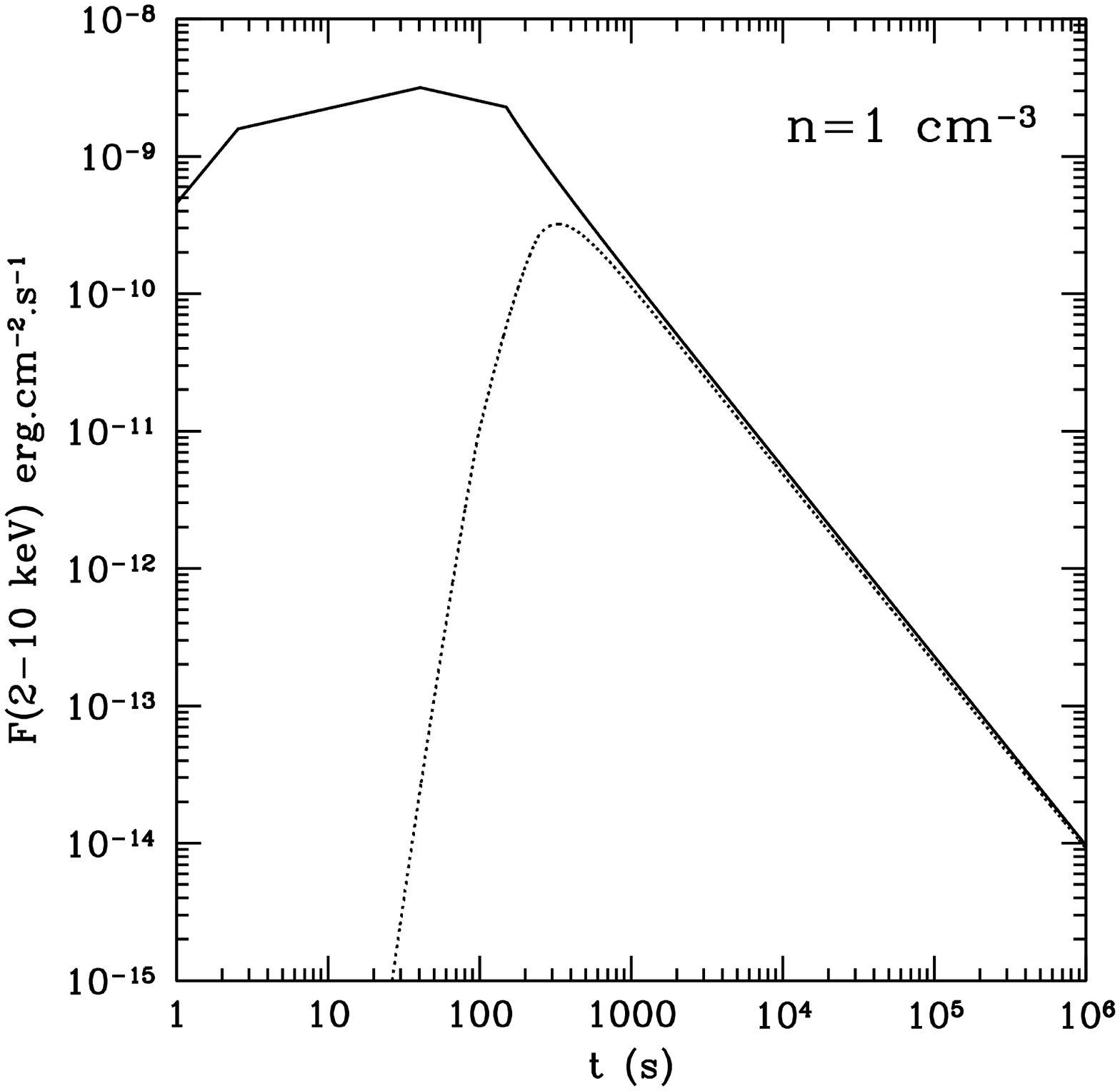}}&
\resizebox{0.48\hsize}{!}{\includegraphics{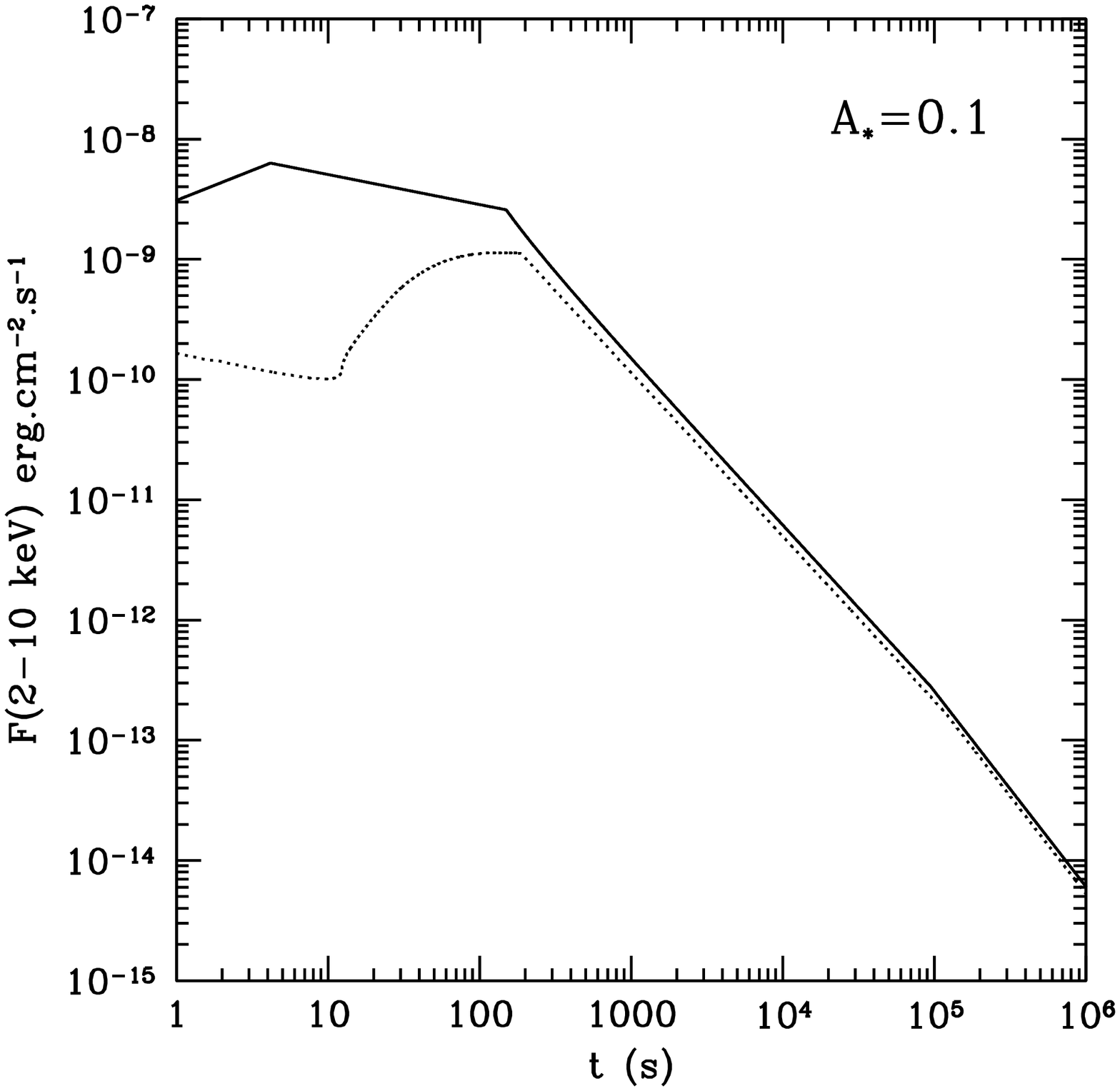}}\\
\resizebox{0.48\hsize}{!}{\includegraphics{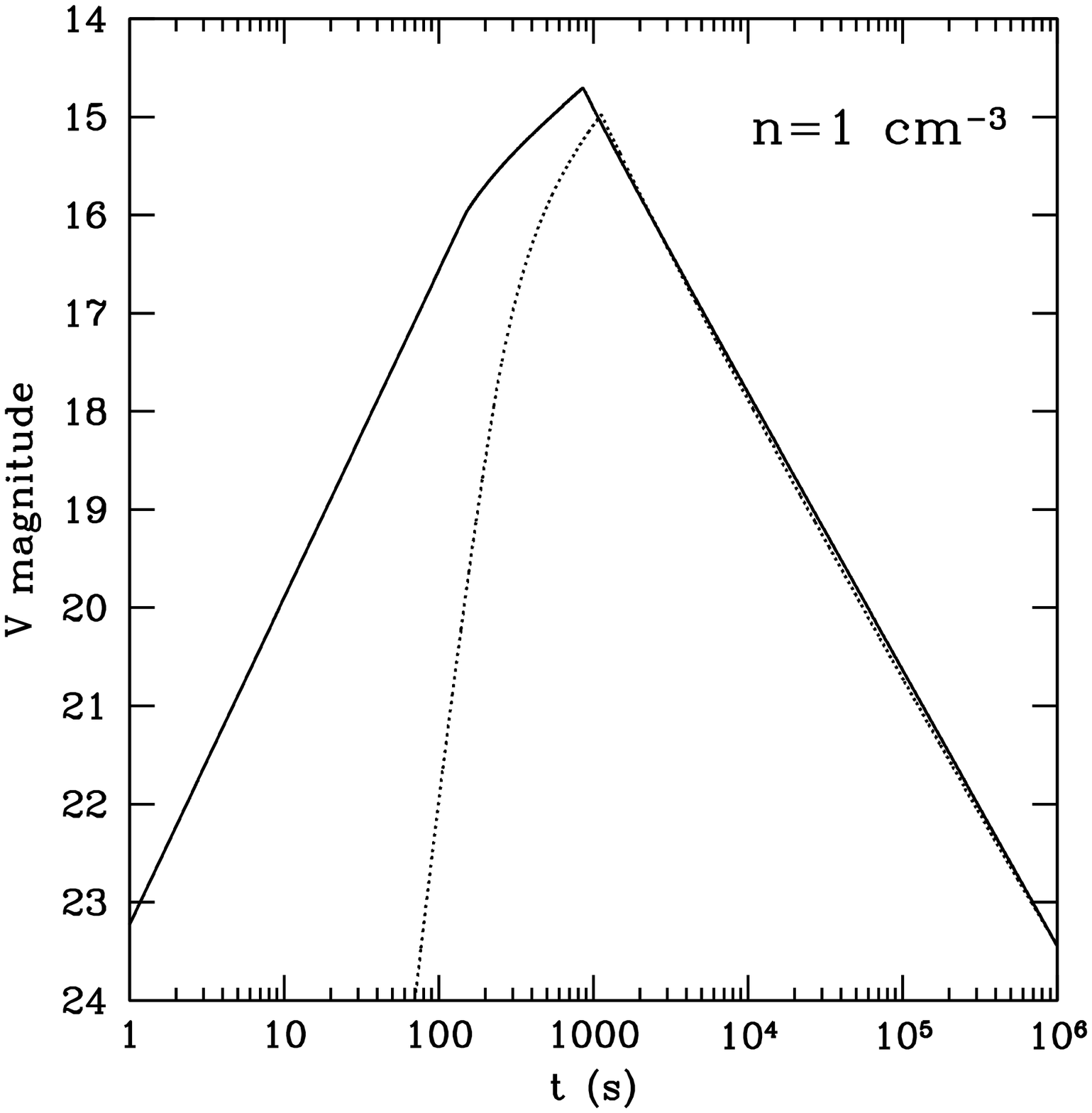}}&
\resizebox{0.48\hsize}{!}{\includegraphics{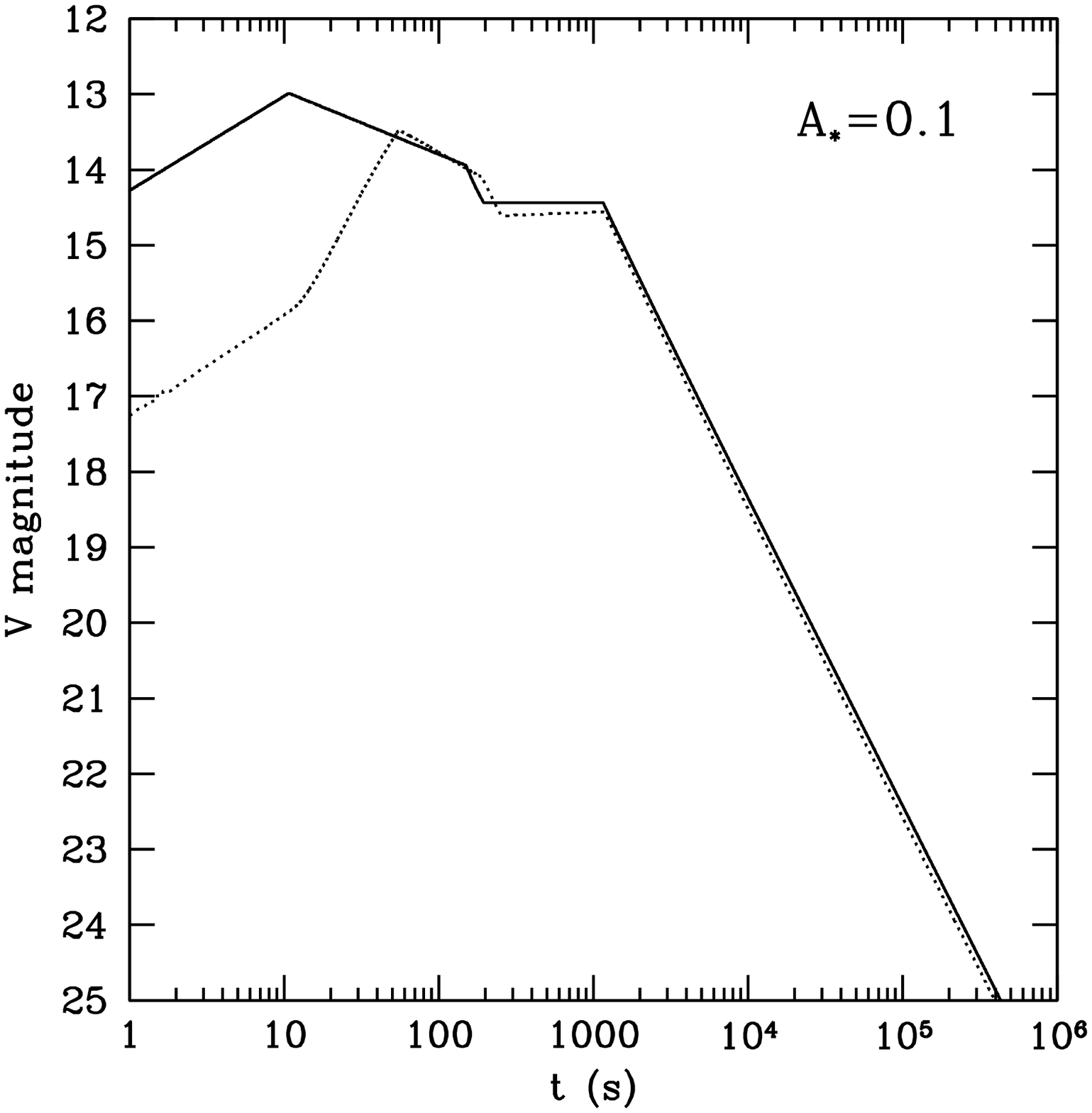}}\\
\end{tabular}
\end{center}
\caption{Afterglow comparison between the EMM (solid line) and the standard 
model (dotted line). Top: X-ray band (2 - 10 keV); Bottom: V band;
Left column: uniform medium of density $n=1$ cm$^{-3}$; 
right column: stellar wind with $A_*=0.1$. A redshift $z=2$ has been 
assumed.}
\end{figure*}
\subsection{Results}
We have represented in Fig.3 afterglow light curves in X-rays
(2 - 10 keV) and in the V band for a uniform external medium and
a stellar wind. The burst parameters 
are identical to those used 
in Fig.2. For comparison we also show in Fig.3 standard afterglows 
computed 
in the internal/external shock model with the same total injected energy. 
A redshift $z=2$, typical of SWIFT bursts has been adopted (Jakobsson et al.,
2006).
In all cases the afterglows are brighter at early times in the EMM 
while at late times (in the Blandford-McKee regime) the models coincide.
The difference in early evolution is larger for a uniform external medium,
especially in X-rays. This is due to the large initial Lorentz factor in the
EMM compared to the standard model (see Fig.1) which leads to a higher
electron Lorentz factor in the shocked external medium.
The frequency $\nu_m$ is then in the X-ray band already at very early
times 
while it is still in the UV/visible in the standard model. 
In the wind case, the differences between the EMM and the standard case
are smaller and the two models will be therefore  
more difficult to distinguish.

For the EMM we have also calculated afterglows when the source is variable
(with $L_{\rm EM}(t)$ given by Eq.(13)). The results are shown in Fig.4 
for a uniform external 
medium and the first 200
seconds of evolution (the source being active for 150 s in the observer frame).
The left panel in Fig.4 shows the simple calculation which only includes 
line of sight emission. It produces
a highly variable X-ray light curve but off-axis effects (time delays
and spectral softening) will smear out any variability occurring on a 
time scale shorter than 
\begin{equation}
\Delta t={R\over 2 c \Gamma^2}
\end{equation}
where $R$ and $\Gamma$ are the radius and Lorentz factor of the emitting shell.
The right panel of Fig.4 shows $\Delta t$ as a function of observer time
together with the X-ray light curve now obtained with a detailed 
calculation including off-axis effects (Granot, Piran \& Sari, 1999;
Woods \& Loeb, 1999). Except for the first two or three peaks 
which partially subsist because $\Delta t$ is initially small, 
the rest of the curve
becomes nearly monotonic.   
In the visible the variability is barely visible, even without including 
off-axis effects. 
\begin{figure*}{}
\begin{center}
\begin{tabular}{cc}
\resizebox{0.48\hsize}{!}{\includegraphics{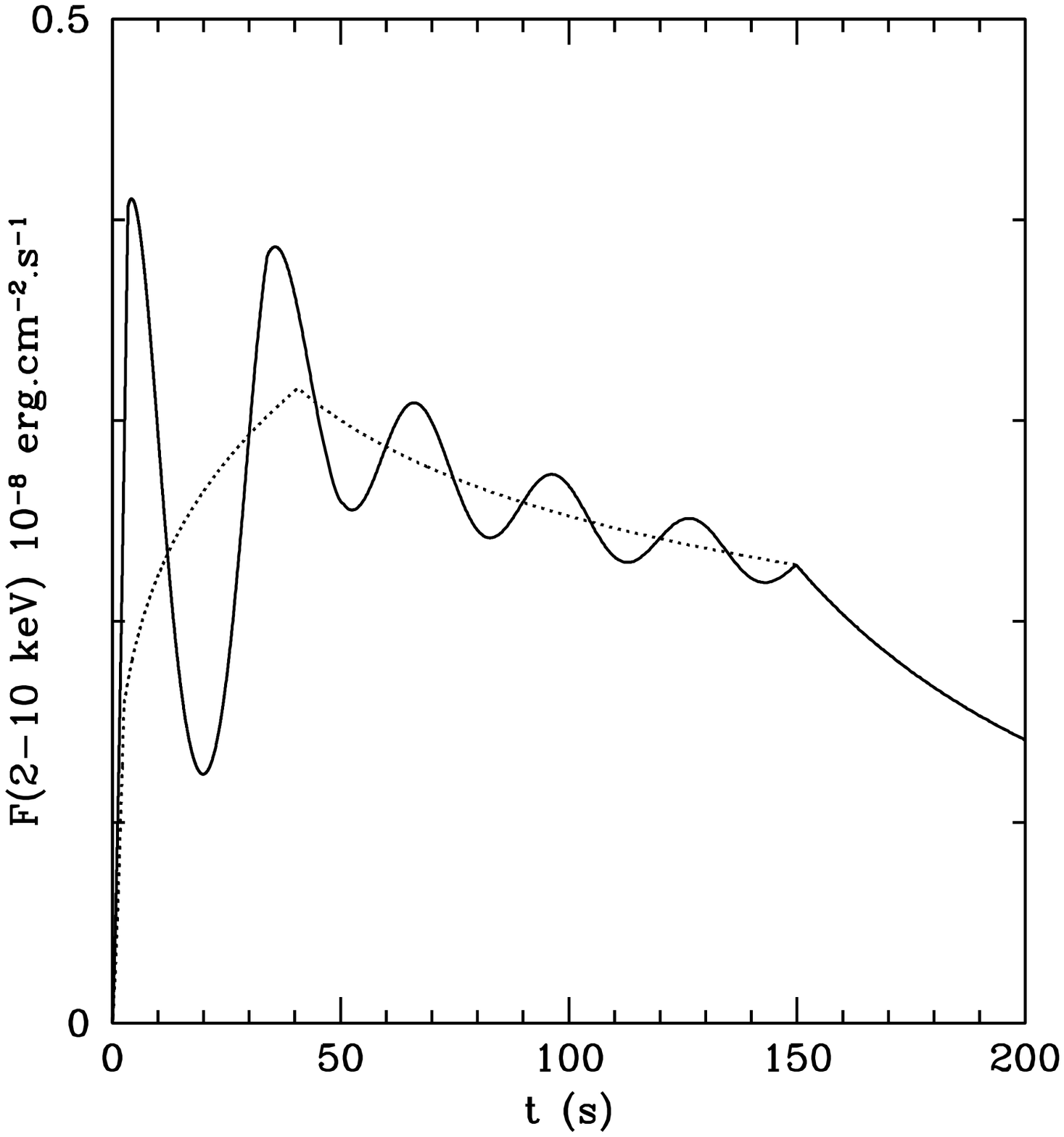}}&
\resizebox{0.48\hsize}{!}{\includegraphics{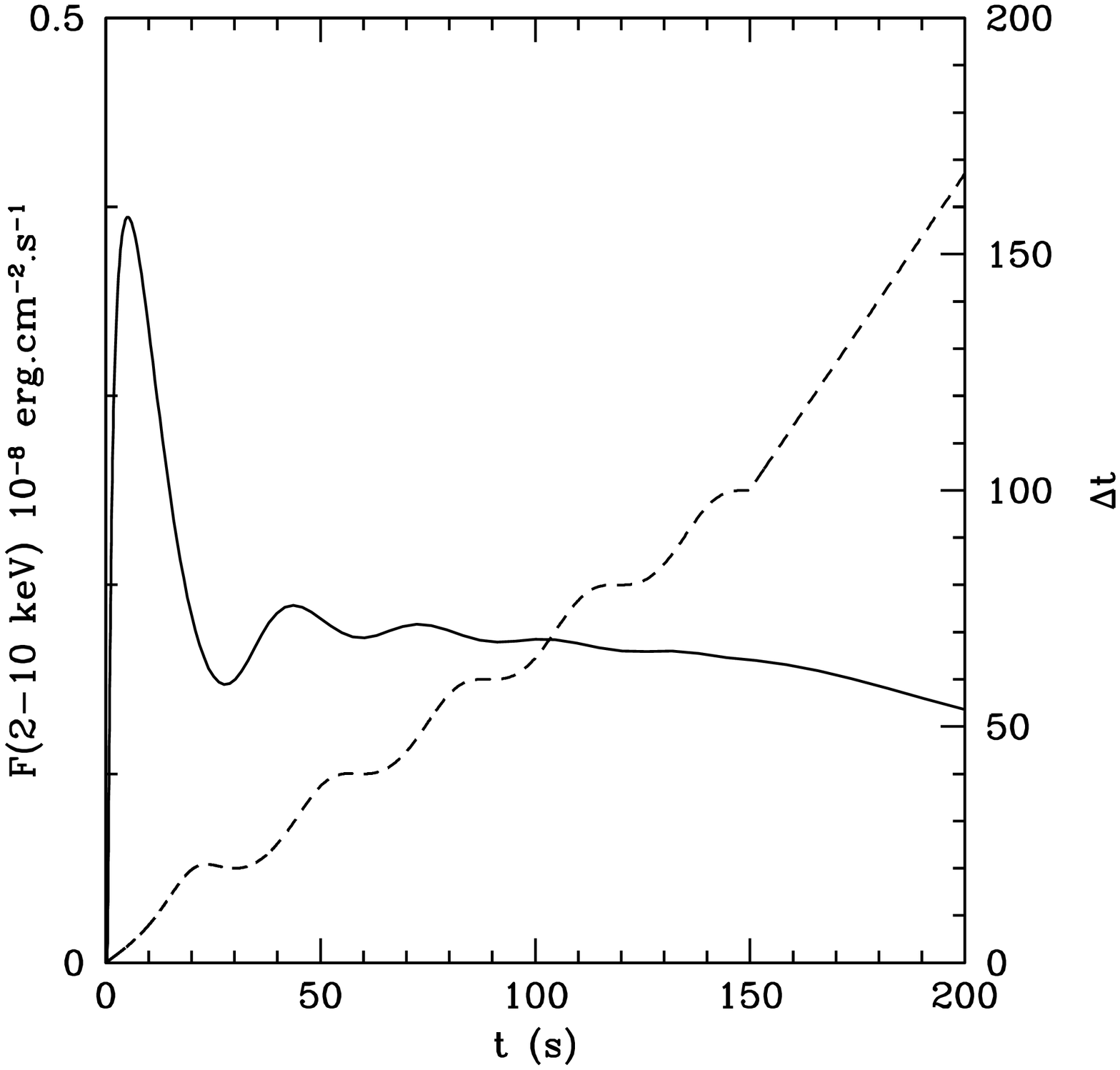}}\\
\end{tabular}
\end{center}
\caption{Early afterglows for a variable source ($L_{\rm EM}(t)$ being given 
by Eq.(13)). Left panel: afterglow computed with on-axis emission only 
compared
to the case of a constant source (dotted line). 
Right panel:  afterglow computed with off-axis emission included
(full line) and 
geometrical delay $\Delta t=R/2c \Gamma^2$ (dashed line).
A uniform external medium
of density $n=1$ cm$^{-3}$ and a redshift $z=2$ have been assumed.}
\end{figure*}

\section{Discussion and conclusion}
The lightcurves in Fig.3 show that the EMM and the standard 
model notably differ at early times (during the
period of source acti\-vi\-ty). 
However relying on these differences
alone to identify the physical origin of GRBs will be a difficult task
requiring a very early follow-up of the afterglow.
In X-rays, SWIFT should be able to do that (at least in some cases) 
but the problem here will come from the
mixing of the afterglow contribution with the brighter prompt emission component. 
This mixing 
will also probably prevent an unambiguous detection of the
imprint of source variability
on the X-ray afterglow (Fig.4). In the visible, where the burst prompt 
emission is weak and probably negligible (see however the recent RAPTOR
observations of GRB 041219a and GRB 050820a (Vestrand et al., 2005, 2006)), the EMM 
predicts a brighter afterglow for a given
set of parameters $\epsilon_e$, $\epsilon_B$, $n$ or $A_*$. But in real
afterglows these parameters are not known a priori and deciding between 
models will be tricky. Polarization properties of the burst prompt emission
(Lyutikov, 2004) when they become more easily accessible may provide clearer  
evidence.  

A last interesting point concerning the EMM is related to the shallow
part observed by SWIFT in many X-ray afterglows. The light curves in
Fig. 3 show that the EMM indeed predicts an initially flat region in 
the early X-ray afterglow. However this flat region does not last more
than the period of source activity (150 s in observer time in Fig.3). It would then extend to $10^4$ s (or more)
only if the source can remain active for that duration, 
as was also
suggested for the
standard model (Zhang et al., 2005).


\end{document}